\documentclass[aps,prl,a4paper,showpacs,twocolumn,floatfix]{revtex4}
\usepackage[latin1]{inputenc}
\usepackage{amsfonts}
\usepackage{amssymb}
\usepackage{amsmath}
\usepackage{calc}
\usepackage{graphicx}
\usepackage{epsfig}
\usepackage{bm}
\usepackage{ulem}
\usepackage{setspace}
\usepackage{subfigure}

\newcommand{\ket}[1]{|#1\rangle}
\newcommand{\bra}[1]{\langle#1|}

\def\text#1{\textrm{#1}}
\def\mn#1{\langle #1 \rangle}

\begin{document}

\title{Quantum experiments with human eyes as detectors based on cloning
via stimulated emission}
\author{Pavel Sekatski$^1$}
\author{Nicolas Brunner$^{1,2}$}
\author{Cyril Branciard$^1$}
\author{Nicolas Gisin$^1$}
\author{Christoph Simon$^1$}
\affiliation{$^1$Group of Applied Physics, University of
Geneva, Switzerland\\$^2$H.H. Wills Physics Laboratory,
University of Bristol, United Kingdom}

\date{\today}

\begin{abstract}
We show theoretically that the multi-photon states obtained
by cloning single-photon qubits via stimulated emission can
be distinguished with the naked human eye with high
efficiency and fidelity. Focusing on the ``micro-macro''
situation realized in a recent experiment [F. De Martini,
F. Sciarrino, and C. Vitelli, Phys. Rev. Lett. {\bf 100},
253601 (2008)], where one photon from an original entangled
pair is detected directly, whereas the other one is greatly
amplified, we show that performing a Bell experiment with
human-eye detectors for the amplified photon appears
realistic, even when losses are taken into account. The
great robustness of these results under photon loss leads
to an apparent paradox, which we resolve by noting that the
Bell violation proves the existence of entanglement {\it
before} the amplification process. However, we also prove
that there is genuine micro-macro entanglement even for
high loss.
\end{abstract}

\maketitle

The basic principles of quantum physics such as quantum
superpositions and entanglement have already had a major
impact on the scientific world view. These phenomena are
typically far removed from our everyday experience. It is
of interest to explore various ways of bringing quantum
phenomena closer to the macroscopic level, and to everyday
life. One possible approach is to ask whether it might be
possible to perform quantum optics experiments with human
eyes as detectors \cite{BBG}. Quantum cloning of single
photon states via stimulated emission
\cite{SWZ,DeM00,Lamas,Fasel,Nagali} has recently allowed
the experimental creation of tens of thousands of clones
starting from a single photon \cite{DeM08}. Here we show
that cloning by stimulated emission is a very promising
approach for the realization of quantum experiments with
human-eye detectors.

The photon detection characteristics of the human eye have
been studied in significant detail starting with Ref.
\cite{pirenne}. Our results are based on the following
theoretical model which describes the experimental evidence
very well \cite{riekebaylor}. The eye is modeled as an {\it
ideal threshold detector preceded by very significant
losses}. More formally, we define the positive operator
corresponding to a detection by the eye as $\hat{E_y}=
C_L^\dag \hat{T_y}C_L$ where
$\hat{T_y}=\openone-\hat{T_n}=\openone-
\sum_{m=0}^{\theta-1} \ket{m} \! \bra{m}$,
%\label{threshold}
%\end{equation}
with photon number states $|m\rangle$, is the projection
operator corresponding to an ideal threshold detector with
threshold $\theta$, and $C_L =e^{\gamma (a^\dag c - a
c^\dag)}\ket{0}_c$
%\label{loss}
%\end{equation}
is the loss channel, where $a$ is the mode that we are
interested in detecting and $c$ is the initially empty mode
whose coupling to $a$ is responsible for the loss. We have
introduced the subscript $y$ to mean ``yes'', corresponding
to a successful detection. Analogously, the operator for a
non-detection is $\hat{E}_n=C_L^\dag \hat{T_n}C_L$. Based
on Ref. \cite{riekebaylor} we choose the values $\theta=7$
for the threshold and $\eta=\cos^2 \gamma=0.08$ for the
transmission of the eye. These values provide an excellent
fit for the experimental response curve of the eye, which
looks like a smoothed out step function, where the step
occurs in the vicinity of ca. 100 photons impinging on the
eye, cf. Fig. 2 of Ref. \cite{riekebaylor}.

It is a priori not easy to design quantum experiments using
the eye as a detector. For example, the approach studied in
Ref. \cite{BBG} of observing large numbers of independent
entangled pairs does not allow the violation of a Bell
inequality if the above realistic eye model is used
\cite{lowvis}, rather than the more idealized model
considered in Ref. \cite{BBG}. Nevertheless, in the present
work we show that quantum experiments with human-eye
detectors become a realistic possibility, if detection with
the naked eye is combined with cloning via stimulated
emission.

Cloning by stimulated emission was originally introduced
\cite{SWZ,DeM00} in the context of {\it universal} cloning
\cite{universal}, i.e. in a setting where all input states
are treated equally. Here we focus instead on {\it
phase-covariant} cloning \cite{phasecov} by stimulated
emission \cite{Nagali}, in order to stay close to the
experiments of Refs. \cite{DeM08,DeM08-bell}. A
phase-covariant cloner makes good copies only of input
states that lie on a great circle of the Bloch sphere, e.g.
the equator. Considering qubits realized by the
polarization states of single photons in a spatial mode
$\bf{a}$, a phase-covariant cloner can be realized based on
stimulated collinear type-II parametric down-conversion
\cite{Nagali}, where the appropriate Hamiltonian for the
down-conversion process is $H= i \chi a^\dag_H a^\dag_V +
h.c.$, where $\chi$ is proportional to the non-linear
susceptibility of the crystal and to the pump power, and
$a_H$ and $a_V$ are the horizontal and vertical
polarization modes corresponding to the spatial mode
$\bf{a}$. Identifying $a_H$ and $a_V$ with the north and
south poles of the Bloch sphere, one can introduce a basis
of ``equatorial'' modes $a_\phi$ and $a_{\phi \perp}$ via
the relations $a_H=\frac{1}{\sqrt2}e^{i\phi}(a_\phi+i\,
a_{\phi\perp})$, $a_V=\frac{1}{\sqrt2}e^{-i\phi}(a_\phi-i
\, a_{\phi\perp})$. Different choices of the phase $\phi$
correspond to different bases. Rewriting $H$ in terms of
$a_\phi$ and $a_{\phi \perp}$ gives $H= \frac{i \chi}{2}
({a_\phi^\dag}\!\,^2+{a_{\phi\perp}^\dag}\!\!\!\!^2 )
+h.c.$; one can see that $H$ has the same form for any
choice of equatorial basis. This is why the cloning process
is phase covariant. We will assume that a choice of basis
has been made and denote the corresponding equatorial modes
by $a$ and $a_{\perp}$ for compactness of notation.

We now show that the multi-photon states obtained by
cloning single-photon qubits via stimulated emission can be
distinguished with the naked eye with a high probability
for a conclusive result and high fidelity. Consider cloning
the two orthogonal single-photon qubit states
$a^{\dagger}|0,0\rangle=|1,0\rangle$ and
$a^{\dagger}_{\perp}|0,0\rangle=|0,1\rangle$. The time
evolution operator for the cloning process is $e^{-iHt}=U
U_{\perp}$ with $U= e^{\frac{g}{2}({a^\dag}^2-a^2)}$,
$U_\perp=e^{\frac{g}{2}( {a_{\perp}^\dag}^2 -a_\perp^2)}$,
where we have defined the amplification gain $g=\chi t$,
with $t$ the interaction time for the down conversion
process. After the amplification, the qubit states become
\begin{eqnarray}
\ket{\Phi}= U U_\perp
\ket{1,0}=\ket{A_1}\ket{A_0}_{\perp},\nonumber\\
\ket{\Phi_\perp}= U U_\perp
\ket{0,1}=\ket{A_0}\ket{A_1}_{\perp}, \label{Phi}
\end{eqnarray}
where we have introduced the notation $|A_1\rangle=U
|1\rangle, |A_0\rangle=U|0\rangle$, and analogously for the
perpendicular modes. It is easy to show, e.g. by
integrating the equations of motion in the Heisenberg
picture, that $U^\dag a^\dag U = \cosh( g) \,a^\dag +
\sinh( g) \,a$, which allows one to calculate the mean
photon numbers in the two states $|A_0\rangle$ and
$|A_1\rangle$, $\bra{A_1}a^\dag a \ket{A_1} = 3
\sinh^2(g)+1$, and $\bra{A_0}a^\dag a \ket{A_0} =
\sinh^2(g)$.  This shows that stimulating the
down-conversion process with a single photon leads to an
approximate tripling of the resulting output photon number
compared to a vacuum input (for large $g$).

Our proposal for distinguishing $\ket{\Phi}$ and
$\ket{\Phi_\perp}$ using human eyes as detectors, which is
illustrated in Fig. 1, is based on this significant
difference in typical photon numbers between the states
$|A_1\rangle$ and $|A_0\rangle$, in combination with the
fact that the eye is a (smooth) threshold detector. The
amplification gain $g$ can be adjusted in such a way that
$|A_1\rangle$ will give a detection by the eye with high
probability (i.e. it is ``above the threshold''), whereas
$|A_0\rangle$ will not (it is ``below the threshold'').
Under these conditions, separating the two modes $a$ and
$a_\perp$ and directing each of them to one eye
\cite{twopersons}, $\ket{\Phi}$ will mostly give rise to
detections in the eye exposed to mode $a$, whereas
$\ket{\Phi_\perp}$ will mostly give rise to detections in
the eye exposed to mode $a_\perp$.

\begin{figure}
    \center
  \includegraphics[width=0.7 \columnwidth]{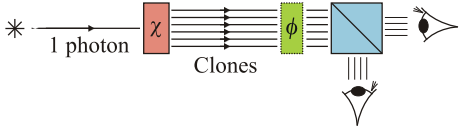}\\
  \caption{A single-photon qubit is
  amplified through cloning via stimulated emission in a non-linear crystal
  ($\chi$). The clones are split
  into two orthogonal polarization modes, and each mode is detected
  by a naked human eye. The polarization basis can be varied with
  the help of a waveplate ($\phi$).}
\end{figure}

Since the eye is not a perfect threshold detector, and
since the photon number distributions in the two states
$|A_0\rangle$ and $|A_1\rangle$ have large variances
\cite{variances}, there will also be events where both eyes
detect something, where none of the eyes detect anything,
or even where only the ``wrong'' eye responds. Introducing
the notation $p(y,n|\Phi)$ for the probability of a
detection (``yes'') in mode $a$ and no detection (``no'')
in mode $a_\perp$, given the state $\ket{\Phi}$, and
analogously for the other cases, one can then define the
probability for a conclusive measurement, corresponding to
a detection in only one eye, as
\begin{equation}
\varepsilon=p(y,n|\Phi)+p(n,y|\Phi)=p(y,n|\Phi_\perp)+p(n,y|\Phi_\perp),
\label{eff}
\end{equation}
where the equality follows from Eq. (\ref{Phi}) (and
$\varepsilon$ stands for ``efficiency''). The accuracy of
the measurement can be quantified via the visibility $V$,
defined as
\begin{equation}
V=\frac{p(y,n|\Phi)-p(n,y|\Phi)}{p(y,n|\Phi)+p(n,y|\Phi)}.
\label{V}
\end{equation}

Based on the above model of the eye as a photon detector,
the probabilities can be expressed as
$p(y,n|\Phi)=\bra{A_1}\hat{E_y}\ket{A_1}\bra{A_0}\hat{E_n}\ket{A_0}$,
and analogous relations for the other probabilities. In
order to evaluate the expectation values of $\hat{E}_y$ and
$\hat{E}_n$, one has to evaluate general terms of the form
$P_{\ket{A_0}}^{\ket{m}}=\bra{A_0} C_L^\dag
\ket{m}\!\bra{m}C_L\ket{A_0}$  and
$P_{\ket{A_1}}^{\ket{m}}=\bra{A_1}
C_L^\dag\ket{m}\!\bra{m}C_L\ket{A_1}$. The projector on a
Fock state $|m\rangle$ can be written as $\ket{m}\! \bra{m}
= \delta_{a^\dag a,m}= \frac{1}{2 \pi} \int_0^{2\pi} dk\,
e^{-ik(a^\dag a-m)}$. The above expressions can be
evaluated using operator ordering techniques that follow
Ref. \cite{Collett}. As a first step one can show that $U
=e^{\frac{1}{2}\tanh \!g\, {a^\dag}^2} e^{-\ln (\cosh
\!g)(a^\dag a + \frac{1}{2})} e^{-\frac{1}{2}\tanh\! g\,
a^2}$ and $C_L= e^{\tan \!\gamma\, a\, c^\dag} e^{\ln (\cos
\gamma)\, a^\dag a}\ket{0}_c$. Furthermore $C_L^\dag e^{-ik
a^\dag a} C_L = e^{\ln(1-\eta + \eta e^{-ik})a^\dag a}
\equiv e^{-\ln(X_0)a^\dag a}$, where we have introduced the
expression $X_0=(1-\eta + \eta e^{-ik})^{-1}$, which allows
us to evaluate $U^\dag C_L^\dag e^{-ik a^\dag a} C_L
U=U^\dag e^{-\ln(X_0)a^\dag
a}U=Y^{-\frac{1}{2}}e^{-\frac{1}{2}\ln X a^\dag
a}e^{\frac{1}{2}Z {a^\dag}^2}e^{\frac{1}{2}Z {a}^2}
e^{-\frac{1}{2}\ln X a^\dag a}$, with $X= X_0\cosh^2 g -
\frac{\sinh^2 g}{X_0}$, $Y=\frac{X}{X_0}$, and
$Z=\frac{1}{2} \partial_g X$. This gives $\bra{A_0}C_L^\dag
e^{-ik a^\dag a} C_L\ket{A_0}= Y^{-\frac{1}{2}}$ and
$\bra{A_1}C_L^\dag e^{-ik a^\dag a} C_L\ket{A_1}=
Y^{-\frac{1}{2}}X^{-1}$, which implies
$P_{\ket{A_0}}^{\ket{m}}= \frac{1}{2\pi}\int dk \, e^{ikm}
Y^{-\frac{1}{2}}=\frac{i}{2\pi}\int_\Gamma dz \,\frac{
Y^{-\frac{1}{2}}}{z^{m+1}}$ and $P_{\ket{A_1}}^{\ket{m}}=
\frac{1}{2\pi}\int dk \, e^{ikm}
Y^{-\frac{1}{2}}X^{-1}=\frac{i}{2\pi}\int_\Gamma dz
\,\frac{ Y^{-\frac{1}{2}}X^{-1}}{z^{m+1}}$, where we made
the change of variable $z = e^{-ik}$ and $\Gamma$ is the
unit circle in the complex plane with clockwise
orientation. Note that $X$ and $Y$ are functions of $z$
through their dependence on $X_0$. It is easy to show that
$X$ and $Y$ are never zero inside $\Gamma$ for finite $g$,
which allows us to apply the Cauchy integral formula,
yielding $P_{\ket{A_0}}^{\ket{m}} =
\frac{1}{m!}\partial_z^m Y^{-\frac{1}{2}}|_{z=0}$, and
$P_{\ket{A_0}}^{\ket{m}} = \frac{1}{m!}\partial_z^m
(Y^{-\frac{1}{2}}X^{-1})|_{z=0}$.

\begin{figure}
    \center
  \includegraphics[width=0.9\columnwidth]{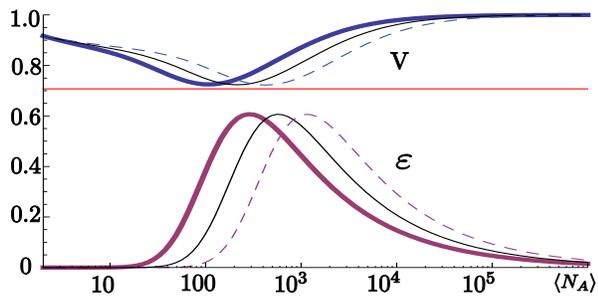}\\
  \caption{Efficiency $\varepsilon$ and visibility $V$, defined in
  Eqs. (\ref{eff}) and (\ref{V}), of the human-eye detection method for
  amplified single-photon qubits, as a function of the mean photon number
  after amplification $\langle N_{\bf{a}} \rangle$ (thick lines). The efficiency has a
  maximum of $\varepsilon=0.61$ for $\langle N_{\bf{a}} \rangle=288$.The visibility
  never drops below $\frac{1}{\sqrt{2}}$, which is relevant for
  Bell experiments in the micro-macro setting of Refs.
  \cite{DeM08,DeM08-bell}, cf. text and Fig. 3.
  We also show $V$ and $\varepsilon$ for the case of additional losses
  after the amplification,
  corresponding to overall transmission factors $\frac{\eta}{2}$ (thin lines)
  and $\frac{\eta}{4}$ (dashed lines).}
  \label{viseff}
\end{figure}

These results make it possible to calculate the detection
probabilities $p(y,n|\Phi)$ etc., and thus the visibility
$V$ and the efficiency $\varepsilon$, as a function of the
gain $g$, which directly determines the mean photon number
after amplification, summed over both polarization modes,
$\langle N_{\bf{a}} \rangle=4 \sinh^2(g)+1$, cf. above. The
results are shown in Fig. \ref{viseff}. One sees that
$\varepsilon$ has a maximum for ca. 300 photons. Despite a
dip in the region of high efficiency, $V$ always stays
greater than $\frac{1}{\sqrt{2}}$, which is an important
bound for Bell experiments, cf. below. We thus see that the
states $|\Phi\rangle$ and $|\Phi_\perp\rangle$ can be
distinguished with high efficiency and accuracy at the same
time. Thanks to the discussed phase covariance, this result
is true for any equatorial basis. Fig. 2 also shows the
effect of other losses after the amplification in addition
to the unavoidable losses in the eye. Since the model of
the eye used is an ideal threshold detector preceded by
losses, this can be done simply by varying the value of
$\eta$. One sees that the effect of losses can be
completely compensated by increasing the gain.

Let us now apply these results to one particular
interesting experimental situation, namely the micro-macro
scenario of Refs. \cite{DeM08,DeM08-bell}, see also Fig. 3.
In these experiments, a first low-gain down-conversion
process creates an entangled photon pair into the two
distinct spatial modes $\bf{a}$ and $\bf{b}$ in a
polarization singlet state,
$|\psi_-\rangle=\frac{1}{\sqrt{2}}(a^{\dagger}_H
b^{\dagger}_V-a^{\dagger}_V b^{\dagger}_H)|0,0,0,0\rangle$,
where $|0,0,0,0\rangle$ denotes the vacuum for all
participating modes. Thanks to the rotational invariance of
the singlet, this can be rewritten in an equatorial mode
basis as $|\psi_-\rangle=\frac{1}{\sqrt{2}}(a^{\dagger}
b^{\dagger}_\perp-a^{\dagger}_\perp
b^{\dagger})|0,0,0,0\rangle$. The photon in the $\bf{b}$
spatial mode is detected directly, whereas the photon in
the $\bf{a}$ mode is greatly amplified with the
phase-covariant cloning process described above, leading to
a micro-macro entangled state
\begin{equation}
|\Psi_-\rangle=\frac{1}{\sqrt{2}}(|\Phi\rangle_{\bf{a}}
|0,1\rangle_{\bf{b}} -|\Phi_\perp\rangle_{\bf{a}}
|1,0\rangle_{\bf{b}}) \label{Psi-}
\end{equation}
(still written in the equatorial basis for both spatial
modes). The capability of human-eye detectors to
distinguish the two states $|\Phi\rangle$ and
$|\Phi_\perp\rangle$ with high visibility implies the
possibility of observing a violation of the CHSH Bell
inequality with the same visibility for this entangled
state, provided that the detection of the un-amplified
photon in mode $\bf{b}$ does not introduce any errors. Note
that measurements in two different equatorial bases for
both systems $\bf{a}$ and $\bf{b}$ are sufficient for
testing the CHSH inequality. The detection of the
un-amplified photon also serves as a trigger, signaling
that a pair has indeed been produced in the low-gain
down-conversion.

\begin{figure}
    \center
  \includegraphics[width=\columnwidth]{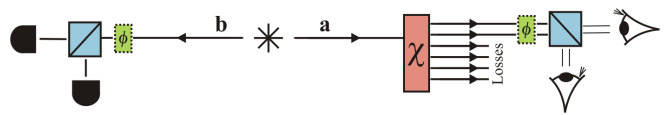}\\
  \caption{We consider the micro-macro entanglement scenario of Refs.
  \cite{DeM08,DeM08-bell}, but with human-eye detectors for
  the macro system.}
  \label{micmac}
\end{figure}

It is worth noting that the proposed measurement by human
eye, although clearly motivated by conceptual rather than
practical considerations, can in fact be orders of
magnitude more efficient than the ``orthogonality filter''
technique used in Refs. \cite{DeM08,DeM08-bell}, for which
the success probability is of order $10^{-4}$. This is
because the eye as a threshold detector is extremely well
suited to the task of discriminating the states
$|\Phi\rangle$ and $|\Phi_\perp\rangle$.

The robustness of the visibility with respect to losses
shown in Fig. 2 means that a strong Bell inequality
violation could be achieved for arbitrarily high losses,
provided that the amplification is sufficiently strong.
This is paradoxical at first sight, since losses are
clearly going to affect the micro-macro entanglement, as
information about the macro-state ($|\Phi\rangle$ or
$|\Phi_\perp\rangle$) leaks into the environment. Even in
the case where there are only the losses intrinsic to the
eye, i.e. for $\eta=0.08$, most of the photons are lost,
such that the environment contains almost all the available
information, which means that the remaining micro-macro
entanglement must be quite small. So how can the visibility
of the Bell violation remain so high?

This apparent paradox can be resolved by realizing that,
while the efficiency of the proposed detection method is
quite high, it is always significantly smaller than one,
such that the measurement is nevertheless post-selective.
Moreover, whereas in the lossless case the macro-system
lives in a two-dimensional Hilbert space spanned by
$|\Phi\rangle$ and $|\Phi_\perp\rangle$, in the presence of
losses it lives in a much larger (in fact, in principle
infinite-dimensional) space. Together these two facts open
up an important ``loophole''. Conclusive (i.e. $(y,n)$ or
$(n,y)$) results for different equatorial bases correspond
to different, almost orthogonal, subspaces of the
high-dimensional Hilbert space. It is not difficult to
construct separable multi-photon states that exploit this
loophole to achieve the same visibility as in Fig.
\ref{viseff} \cite{sepmodel}. The experimental observation
of such a visibility by itself therefore allows no
conclusion about the existence of micro-macro entanglement.

Nevertheless, the same measurements do allow one to prove
the {\it entanglement of the original entangled pair}
before amplification. From this perspective, the
amplification and losses can be simply seen as part of the
detection process for the original single photon. The
Hilbert space of the original photon is only
two-dimensional, so there is no risk of different subspaces
being detected for different choices of measurement basis.
Moreover, the detection efficiency is independent of the
choice of equatorial basis thanks to the phase covariance
of the amplification. For proving non-locality (as opposed
to just entanglement), there is still the usual detection
loophole due to the limited measurement efficiency.
However, it is no more severe than for any other detection
method that has comparable efficiency. Let us note that the
amplification-based detection is different from
conventional photon detection in one interesting way.
Namely, in the present scenario the choice of detection
basis can be made after the amplification process. This is
quite different compared to conventional discussions of the
measurement process, where the amplification only occurs
after the choice of basis.

Briefly relaxing our focus on human eyes as detectors, we
now show that proving {\it genuine micro-macro entanglement
in the presence of losses} is possible using measurements
that are not post-selective. Using the same methods as in
Ref. \cite{SB} one can derive the following condition,
which has to be fulfilled for all separable states:
$|\mn{\vec{J_{\bf{a}}}\cdot \vec{J_{\bf{b}}}}| \leq
\mn{N_{\bf{a}} N_{\bf{b}}}$. As in Ref. \cite{SB},
$\vec{J_{\bf{a}}}$ and $\vec{J_{\bf{b}}}$ are the Stokes
(polarization) vectors corresponding to two different
spatial modes of the light field, and $N_{\bf{a}}$ and
$N_{\bf{b}}$ are the corresponding photon number operators.
In particular, one can choose a convention where
$J_{z{\bf{a}}}=a^{\dagger}_H a_H-a^{\dagger}_V a_V$, and
$J_{x{\bf{a}}}=a^{\dagger} a-a^{\dagger}_\perp a_\perp$,
i.e. the $x$ direction is identified with the arbitrary
phase choice $\phi$ that was used to define the modes $a$
and $a_\perp$ above. Let us emphasize that the dynamics of
$J_y$ (in fact, of any Stokes vector component in the $x-y$
plane) will be exactly equivalent to that of $J_x$. For our
micro-macro scenario, the state of ${\bf{b}}$ is a
single-photon state, leading to the simplified criterion
$|\mn{\vec{J_{\bf{a}}}\cdot \vec{\sigma_{\bf{b}}}}| \leq
\mn{N_{\bf{a}}}$, where $\vec{\sigma_{\bf{b}}}$ is the
vector of Pauli spin matrices. This means that we have to
evaluate in particular $\mn{\vec{J_{\bf{a}}}\cdot
\vec{\sigma_{\bf{b}}}}= \bra{\Psi_-} C_{L{\bf{a}}}^\dag
\vec{J_{\bf{a}}} \cdot \vec{\sigma_{\bf{b}}}C_{L{\bf{a}}}
\ket{\Psi_-}$ for $|\Psi_-\rangle$ from Eq. (\ref{Psi-}).
One can show quite easily that $\bra{\Psi_-}
C_{L{\bf{a}}}^\dag J_{z{\bf{a}}} \sigma_{z{\bf{b}}}
C_{L{\bf{a}}} \ket{\Psi_-}=\eta$, whereas for the
equatorial components $\bra{\Psi_-} C_{L{\bf{a}}}^\dag
J_{x{\bf{a}}} \sigma_{x{\bf{b}}} C_{L{\bf{a}}}
\ket{\Psi_-}=\bra{\Psi_-} C_{L{\bf{a}}}^\dag J_{y{\bf{a}}}
\sigma_{y{\bf{b}}} C_{L{\bf{a}}} \ket{\Psi_-}=\eta
(\bra{A_1}a^\dag a \ket{A_1}- \bra{A_0}a^\dag a
\ket{A_0})=\eta(2 \sinh^2 g+1)$. On the other hand,
$\langle N_{\bf{a}} \rangle=\bra{\Psi_-} C_{L{\bf{a}}}^\dag
(a^\dag a + a_\perp^\dag a_\perp)C_{L{\bf{a}}} \ket{\Psi_-}
= \eta (\bra{A_1}a^\dag a \ket{A_1}+ \bra{A_0}a^\dag a
\ket{A_0})=\eta (4 \sinh^2 g+1)$, which finally yields
$|\mn{\vec{J_{\bf{a}}}\cdot
\vec{\sigma_{\bf{b}}}}|-\mn{N_{\bf{a}}}= 2 \eta$. One can
see that the violation of this genuine micro-macro
entanglement criterion is sensitive to photon loss as
expected. However, some micro-macro entanglement persists
even for high loss. Note that experimentally demonstrating
micro-macro entanglement in this way would require counting
large photon numbers with single-photon accuracy. As a
consequence of the above-mentioned ``loophole'', we are not
aware of a way of demonstrating genuine micro-macro
entanglement in the presence of losses with human eye
detectors, or with the orthogonality filter technique of
Refs. \cite{DeM08,DeM08-bell}.

We have shown that quantum experiments with human eyes as
detectors appear possible, based on a realistic model of
the eye as a photon detector. We note that these results
remain valid even if losses not only after, but also before
and during the amplification process, are taken into
account, which is possible using similar techniques as in
the present paper \cite{Sekatski}. Motivated by recent
experiments \cite{DeM08,DeM08-bell}, we focused on a
micro-macro scenario, but other experiments, such as
bunching of amplified single photons, or macro-macro
experiments where both photons from an original pair are
amplified, can also be considered. The latter scenario
requires a heralded source of photon pairs, since otherwise
the amplified vacuum, which is invariant under equatorial
rotations, will dominate all detections. Moreover universal
cloning can be considered instead of phase-covariant
cloning. The proposed experiments will require quantum
amplifiers that operate at visible wavelenghts, and pulse
durations that are adapted to the timescales of the human
eye. We intend to address all these points in more detail
in a future publication \cite{Sekatski}.

This work was supported by the Swiss NCCR {\it Quantum
Photonics} and by the EU Integrated Project {\it Qubit
Applications}. We thank M.J. Collett, F. De Martini, S.
Gonzalez-Andino and A. Lvovsky for helpful comments and
useful discussions.


\begin{thebibliography}{99}
\bibitem{lowvis} In fact, one can show that the achievable
visibility of the correlations for this approach is only of
order 0.03, whereas a visibility of
$\frac{1}{\sqrt{2}}=0.71$ is required for violating the
CHSH Bell inequality. This will be shown in more detail in
Ref. \cite{Sekatski}.
\bibitem{twopersons} Note that the eyes should not
belong to the same person, since a single observer is
usually not aware which of his eyes sees a particular light
signal.
\bibitem{variances} The variances are $6 \sinh^4
g+ 4 \sinh^2 g$ for $|A_1\rangle$ and $2 \sinh^4 g+2
\sinh^2 g$ for $|A_0\rangle$.
\bibitem{sepmodel} For example, the following separable
$N+1$ photon state achieves a violation of the Bell
inequality with post-selection:
$\rho=\frac{1}{2\pi}\int_0^{2\pi} d\phi U(\phi) |N,0\rangle
|0,1\rangle \langle N,0|\langle 0,1| U(\phi)^{\dagger}$,
where $U(\phi)$ is a rotation of the whole system around
the $z$ axis by an angle $\phi$. The same model also
achieves high visibility for the orthogonality filter
measurements of Refs. \cite{DeM08,DeM08-bell}. These points
will be discussed in more detail in Ref. \cite{Sekatski}.
\end{thebibliography}
\end{document}